\documentclass{aastex63}
\usepackage{color,soul}

\received{January 15, 2021}
\revised{February 21, 2021}
\accepted{February 23, 2021}
\shorttitle{An Unexpected Outburst of the $\alpha$ Carinid Meteor Shower}
\shortauthors{Bruzzone et al.}
\usepackage{changepage}
\begin{document}
\title{Observation of the $\alpha$ Carinid meteor shower 2020 unexpected outburst.}
\correspondingauthor{J.S. Bruzzone}
\email{juan.s.bruzzone@nasa.gov}

\author[0000-0002-2731-0397]{Juan Sebastian Bruzzone}
\affiliation{Department of Physics, Catholic University of America, 620 Michigan Ave., N.E. Washington, DC 20064, USA}
\affiliation{ITM Physics Laboratory,NASA/Goddard Space Flight Center, 8800 Greenbelt Rd., Greenbelt MD 20771, USA}

\author[0000-0002-0439-9341]{Robert J. Weryk}
\affiliation{Institute for Astronomy, University of Hawaii, 2680 Woodlawn Drive, Honolulu HI 96822, USA}

\affiliation{ITM Physics Laboratory,NASA/Goddard Space Flight Center, 8800 Greenbelt Rd., Greenbelt MD 20771, USA}

\author[0000-0001-8615-5166]{Diego Janches}
\affiliation{ITM Physics Laboratory,NASA/Goddard Space Flight Center, 8800 Greenbelt Rd., Greenbelt MD 20771, USA}

\author[0000-0001-7104-5992]{Carsten Baumann}
\affiliation{Deutsches Zentrum f\"ur Luft- und Raumfahrt, Institut f\"ur Solar-Terrestrische Physik, Neustrelitz, Germany}

\author[0000-0002-7909-6345]{Gunter Stober}
\affiliation{Institute of Applied Physics \& Oeschger Center for Climate Change Research, Microwave Physics, University of Bern, Bern, Switzerland}

\author[0000-0003-4533-3282]{Jose Luis Hormaechea}
\affiliation{Facultad de Ciencias Astronomicas y Geof\'isicas, Universidad Nacional de La Plata, Argentina}
\affiliation{Estaci\'on Astronomica Rio Grande, Rio Grande, Tierra del Fuego, Argentina}
%\linenumbers

\begin{abstract}
\begin{center}
We present observations of the sudden outburst of the $\alpha$ Carinid meteor shower recorded with the Southern Argentina Agile MEteor Radar-Orbital System (SAAMER-OS) near the South Toroidal sporadic region. The outburst peaked between 21 UT and 22 UT on October 14, 2020 and lasted 7 days $(199^{\circ}\leq\lambda_{\odot}\leq 205^{\circ})$ with a mean Sun-centered geocentric ecliptic radiant of $\lambda_{g}-\lambda_{\odot}=271^{\circ}.04$, $\beta_{g}=-76^{\circ}.4$, and a geocentric speed of 33.3 km s$^{-1}$. Assuming a mass index value of $s=2.0$, we compute a peak 24 hour-average flux of 0.029 met. km$^{-2}$ hr$^{-1}$ to a limit of 9th magnitude, which is equivalent to a zenithal hourly rate (ZHR) of 5.7, and comparable to other established showers with similar mass indices. By further estimating the peak fluxes for other typical mass index values, we find that the outburst likely never exceeded a maximum ZHR of $\sim44$, well below the activity of other strong showers. The mean orbital elements resemble those of a short-period object: $a=3.5\pm0.1$ au, $q\simeq 1$ au, $e=0.72\pm0.02$, $i=55^{\circ}.8\pm0^{\circ}.3$, $\omega=1^{\circ}\pm 173^{\circ}$, $\Omega=21^{\circ}.7$, and are similar to those derived for two previous shower outbursts observed with SAAMER-OS at high southern ecliptic latitudes. Using the $D^{\prime}$ criterion did not reveal a parent object associated with this shower in the known object catalogues.
\end{center}
\end{abstract}

\section{Introduction}
The Solar System, like other planetary systems (i.e. $\beta$ Pictoris, \citealt{Burrows95}; Fomalhaut, \citealt{Kalas05}), contains a circumsolar second-generation dusty disk known as the Zodiacal Dust Cloud (ZDC) populated by debris from asteroid collisions and the breakup and activity of comets and interstellar medium grains  (\citealt{Jenniskens2006}, \citealt{Nesvorny2010}). Dedicated radar and optical meteor surveys probe the dust content in the inner solar system via detection of meteoroid ablation high in Earth's atmosphere providing a reliable way to examine the dissemination of material populating the ZDC \citep{Baggaley2002, Brown2010, Janches2015, Jenniskens2011}. Observations of the influx of material at Earth's atmosphere reveal two populations clearly distinguishable in the distribution of the ZDC: the sporadic background sources which comprise of dynamically-evolved sub mm-sized meteoroids and micron-size dust grains largely affected by radiation pressure, and meteor showers, streams of younger and larger meteoroids that in principle could be linked dynamically to parent bodies, namely asteroids and comets, due to their similarity in orbital elements \citep{Jenniskens2006}. 

Surveying, identifying, and studying the meteoroid population is highly relevant, both scientifically and from an operational standpoint. Collisions with $1\mu g-10g$ meteoroids moving at average relative speeds in excess of 30 km s$^{-1}$ could constitute a risk to satellite operations and present a safety concern for manned space missions in low orbit. For example, showers like the Daytime Arietids (ARI), the Geminids (GEM) and the Quadrantids (QUA) attain fluxes for 0.3-cm-equivalent particles near the limits for pressure vessel perforation \citep{Moorhead2019}. All-sky meteor surveys capable of conducting uninterrupted observations and delivering timely reports of sudden changes in meteor activity are thus highly desirable. Furthermore, recording long-term seasonal variations of the sporadic background provide constraints to meteoroid stream models \citep{McNamara2004}. Meteor orbit radars currently in operation like the Canadian Meteor Orbit Radar \citep[CMOR,][]{Webster2004, Brown2008} and the Southern Argentina Agile MEteor Radar Orbital System \citep[SAAMER-OS,][]{Janches2015} aim to address these objectives by continuously monitoring the sub-mm sized meteoroid population ($10^{-7}$ kg$ - 10^{-8}$ kg) to collect a large quantity of meteoroid orbit and flux datasets to characterize meteor showers and sporadic sources \citep{Brown1995, Brown2010, Bruzzone2015, Campbell-Brown2015, Janches2015, Pokorny2014, Pokorny2017}. 

Sudden changes in meteor shower activity are not unusual, and in some cases, may be dramatic. For instance, showers like the Leonids (LEO) or the Draconids (DRA) can produce storms with levels of activity thousand of times higher than normal \citep{Kronk2014}. On the other hand, some shower outbursts can be relatively mild enough to only raise the flux slightly above the radar detection threshold making observations of it possible for the first time. Recent examples of abrupt outbursts include the unexpected Draconid (DRA) meteor storm on October 8, 2012 \citep{Ye2014} and two shower outbursts detected at austral latitudes: the Volantids shower (VOL) outburst ($\beta=-77^{\circ}.7$) in late 2015  \citep{Jenniskens2016, Younger2016, Pokorny2017}, and the $\beta$ Tucanid / $\delta$ Mensid outburst ($\beta=-77^{\circ}.2$) in early 2020 \citep{Janches2020outburst}. \cite{Jenniskens2020ACari} reported significant meteor activity from the weak $\alpha$ Carinid shower on the night of October 13-14, 2020, recording 130 orbits with the Cameras for Allsky Meteor Surveillance \citep[CAMS;][]{Jenniskens2011}.  The $\alpha$ Carinid is a high-latitude southern shower, $\alpha_{g}=103^{\circ}.2$, $\delta_{g}=-57^{\circ}$, $v_{g}=30.1$ km s$^{-1}$, and first reported in \cite{Jenniskens2018} based on 121 orbits with CAMS. In this work, we report the unpredicted outburst of the $\alpha$ Carinid shower as detected by SAAMER-OS, peaking on October 14, 2020 ($\lambda_{\odot}=201^{\circ}$). In Section \ref{sec:radar} we describe the radar and the daily shower monitoring methodology. We provide results and characterize the outburst radiants, orbits and fluxes in Sections \ref{sec:wavrads} and \ref{sec:flux}. Lastly, conclusions and final remarks are presented in Section \ref{sec:conclusions}.

\section{Instrumentation and data analysis}\label{sec:radar}
SAAMER-OS is a VHF all-sky multi-station backscatter meteor orbit radar hosted by the Estaci\'on Astron\'omica Rio Grande (EARG) in Rio Grande, Tierra del Fuego, Argentina \citep{Janches2015}.  Here we present a brief overview of the system and refer the reader to \cite{Janches2020outburst} for a more in-depth review of the hardware and data analysis capabilities. SAAMER-OS is a SKiYMET radar system \citep{Hocking1997}, comprised of a main site (SAAMER-C; $53^{\circ}.786$ S, $67^{\circ}.751$ W) and four remote stations: SAAMER-S ($53^{\circ}.852$ S, $67^{\circ}.76 W$) located at approximately 7 km south of the SAAMER-C; SAAMER-N ($53^{\circ}.682$ S, $67^{\circ}.871$ W) at 13 km northwest of the central station; SAAMER-W ($53^{\circ}.828$ S, $67^{\circ}.842$ W) approximately 7 km southwest of SAAMER-C and SAAMER-E ($53^{\circ}.772$ S, $67^{\circ}.727$ W), at roughly 4 km northeast of main site. The main site hosts the 64 kW (peak power) transmitter with a single three-cross element Yagi transmitting antenna, and the five three-element crossed yagi receiving antenna interferometer array \citep{Hocking1997, Jones1998}. At the main site, meteors are detected as backscatter echoes from meteor trails a few km in length \citep{Kaiser1956} with average interferometric errors less than $0^{\circ}.5$. SAAMER-OS transmits 32.55 MHz pulses with a repetition frequency of 625 Hz, and employs a 7-bit Barker code to achieve a spatial resolution of 1.5 km. Each remote station is equipped with an identical single three-element crossed Yagi receiving antenna to detect the slightly forward scattered signals off the meteor trails. The time delays between the detection of meteors at the main site and the each remote site allows for the determination of the meteoroid speed and its trajectory. SAAMER-OS currently employs an empirical meteor deceleration correction to better estimate the true out-of-atmosphere meteoroid speed \citep{Bruzzone2020}. SAAMER-OS's software suite for event detection, correlation and orbit computation is similar to the one employed in CMOR \citep{Weryk2012} and runs in parallel to the SKiYMET's standard software routines \citep[SKYCORR;][]{Hocking2001}.  Daily detection counts can exceed 10,000 meteoroid orbits \citep{Janches2020outburst}, to a limiting radio magnitude of +9.0, equivalent to meteoroids of mass $10^{-8}$ kg (or 300$\mu$m in diameter) at 30 km s$^{-1}$ \citep{Verniani1973}.

The SAAMER-OS data reduction pipeline employs a 3-D wavelet transform algorithm that is well suited to the daily detection of meteor shower radiants. This method was first used by the Advanced Meteor Orbit Radar \citep[AMOR;][]{Baggaley1994, Galligan2002}, which operated near Christchurch, New Zealand, to probe for clustering of meteor radiants. Since then, it has been applied by other meteor radar surveys \citep{Brown2008, Brown2010, Bruzzone2015, Pokorny2017, Schult2018}, and more recently, it has been applied to radar and optical meteor observations with SAAMER-OS and CAMS in \cite{Bruzzone2020}. We employ the wavelet transform to isolate meteor showers as spatial and temporal enhancements in radiant space of geocentric Sun-centered ecliptic coordinates and geocentric speed: $\left(\lambda_{g}-\lambda_{\odot},\beta_{g}, v_{g}\right)$. Meteoroids that belong to a specific shower concentrate in radiant space and time with a characteristic spread in angular coordinates and speed that differs from the sporadic meteor background. For a given radiant distribution, the wavelet transform returns a list of wavelet coefficients ($W_{c}$) that serve as a metric for clustering in radiant space enhancing the presence of showers. The wavelet transform can be further optimized to amplify the presence of meteor showers by adjusting the wavelet kernel scale dimensions to resemble the shower's natural spread in radiant space \citep{Bruzzone2015}. In this way, meteor showers can be effectively separated from the activity of the sparse sporadic meteor background. For SAAMER-OS, we adopt the wavelet kernel scale parameters $\sigma_{a}=2.5^{\circ}$ and $\sigma_{v}=15\%$ derived in \cite{Pokorny2017}. For the analysis of SAAMER-OS daily observations, the wavelet-based algorithm is evaluated for $\lambda_{g}-\lambda_{\odot}\in \left[0^{\circ},360^{\circ}\right)$ and $\beta_{g}\in\left(-90^{\circ}, 40^{\circ}\right]$ at $0.^{\circ}5$ steps and for $v_{g} \in [10$ km s$^{-1}$, 80 km s$^{-1}$] at 5\% steps, while advancing at $1^{\circ}$ steps in $\lambda_{\odot}$. For each day, this procedure returns a list of $W_{c}$ for which each individual entry is compared to its yearly median and standard deviation. Those entries that exceed 3 times the total standard deviation above the yearly median, $\sigma$, are stored and used in a global maxima search. We proceed to identify a shower core candidate as the radiant returning the maximum in $W_{c}$. Each shower core identified is cross referenced with a compiled list of known meteor showers \citep{Brown2010, Pokorny2017}. When the location of a core candidate is within 3$^{\circ}$ and 15\% in $v_{g}$ of a shower in the reference list, a match is recorded and the candidate is labeled with the shower IAU code in a radiant density map. Radiants in the map are color-coded by the number of adjacent radiants within $2.^{\circ}5$ and serve as a proxy for local enhancements in showers. Those shower core candidates that do not match with any known shower are stored and plotted for review. In this study, we revisit the daily reports for the $\alpha$ Carinid outburst and repeat the wavelet-based analysis at $0.^{\circ}1$ steps in $\left(\lambda_{g}-\lambda_{\odot},\beta_{g}\right)$ and 1.5\% in $v_{g}$ to secure a more precise radiant position and speed. We then follow \cite{Bruzzone2020} to track the outburst progression with time by linking the shower core radiants at each degree in solar longitude.
% The peak transmitting power of SAAMER-OS exce
\section{Results and Discussion}\label{sec:results}
\begin{figure}[ht!]
 \centering
\includegraphics[width=5.0in]{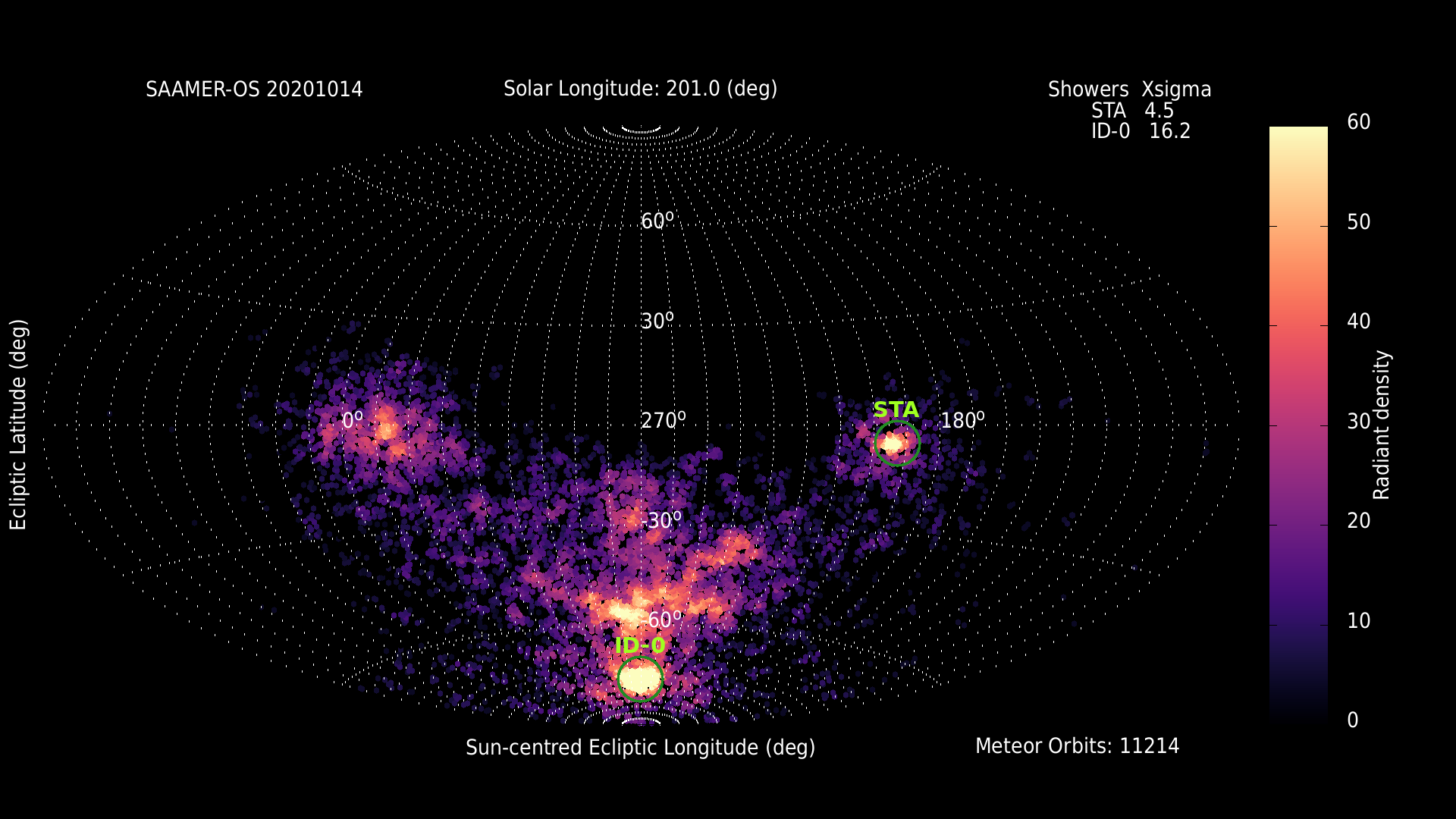}
  \caption{SAAMER-OS daily radiant density plot in Sun-centered geocentric ecliptic coordinates on October 14 2020. The Southern Taurids (STA) is labeled on the map while the $\alpha$ Carinid outburst is labeled as shower candidate ID-0. }\label{fig_map}
\end{figure}
\subsection{Wavelet-based activity profile and orbits}\label{sec:wavrads}
\begin{figure}[ht!]
 \centering
  \includegraphics[width=6.5in]{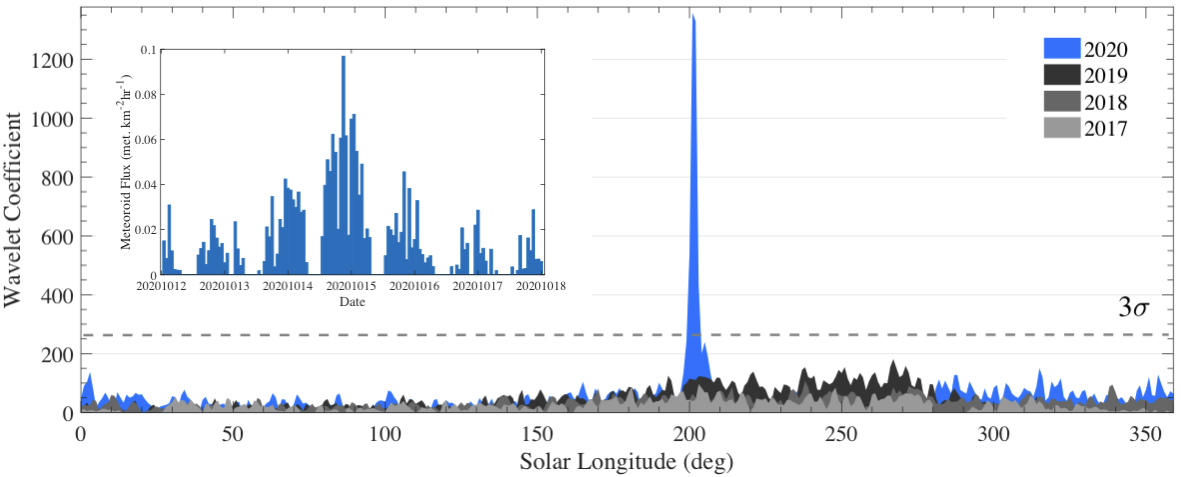}
  \caption{Wavelet coefficient  ($Wc$) profiles at 1$^{\circ}$ steps in $\lambda_{\odot}$ for 2017 through 2020. $W_{c}$ values are estimated at the radiant position and speed of the outburst during the peak at $\lambda_{\odot}=201^{\circ}$. The horizontal dashed line marks the $3\sigma$ level above the annual median $W_{c}$ for 2020. The insert shows the hourly flux from 0 UTC October 12 through 0 UTC October 18, 2020 for a mass index value of 2, times of very low and zero meteor flux indicate reduced radar detectability of meteors when the shower radiant is close to the local zenith.} \label{fig_profile}
\end{figure}

Figure \ref{fig_map} shows the radiant density plot with meteor detections over a 24 hr-period on October 14 2020 in Sun-centered geocentric ecliptic coordinates. The wavelet-based analysis pipeline labels the position of the Southern Taurids as STA, and the $\alpha$ Carinid outburst as candidate ID-0 as it does not match with any known shower in our reference catalog. The procedure returns the radiant location for the outburst at $\lambda_{g}-\lambda_{\odot}=271^{\circ}.04$ ($\alpha_{g}=98^{\circ}.905$), $\beta_{g}=-76^{\circ}.4$ ($\delta_{g}=-53^{\circ}.671$), and $v_{g}=33.3$ km s$^{-1}$ achieving a maximum $W_{c}=1358$ on October 14, 2020  ($\lambda_{\odot}=201^{\circ}$) with a strong detection of $16.2\sigma$ based on 1352 meteors. The wavelet analysis identifies the outburst activity for six consecutive days ($199^{\circ}\leq\lambda_{\odot}< 205^{\circ}$) starting on October 12 through  October 17, 2020. The radiant position and speed measured with SAAMER-OS agree with the mean values from 130 optical meteors reported by \cite{Jenniskens2020ACari} ($\alpha_{g}=98^{\circ}.7$, $\delta_{g}=-54^{\circ}.3$, $v_{g}=32.4$ km s$^{-1}$) with a difference of $0.64^{\circ}$ and 0.9 km s$^{-1}$ respectively. Such a difference in radiant position and speed between SAAMER-OS and CAMS video observations is in agreement with mean values found for a selection of 20 established meteor showers in \cite{Bruzzone2020}. Figure \ref{fig_profile} shows the annual activity profiles by computing $W_{c}$ at $\lambda_{g}-\lambda_{\odot}=271.04^{\circ}$, $\beta_{g}=-76.4^{\circ}$ and $v_{g}=33.3$ km s$^{-1} $ while advancing at $1^{\circ}$ steps in $\lambda_{\odot}$ for 2017 through 2020. A horizontal dashed line indicates the $3\sigma$ level above the median $W_{c}$ of  51.4 for 2020. The profiles confirm the absence of this shower in past years with SAAMER-OS data, and its sudden appearance in 2020. The insert in Figure \ref{fig_profile} displays the hourly meteoroid flux between 0 UT October 12 and 0 UT October 18 2020. Flux estimates with SAAMER-OS are corrected for observational biases by adjusting the observed meteor rates by the radar response function \citep[RRF,][]{Ceplecha1998, Galligan2004} and the variation of the radar effective collecting area with time. Meteor rates are estimated using a 3$^{\circ}$-radius aperture centered in the outburst radiant positions returned by the wavelet transform. We measure a peak hourly 9 mag-flux of 0.097 met. km$^{-2}$ hr$^{-1}$ down to a limiting mass of $1.9\times10^{-8}$ kg. The peak activity ranged between 21 UTC and 22 UTC on October 14, $\lambda_{\odot}\simeq 201.7^{\circ}$, roughly $1^{\circ}$ apart of the time of peak activity at $\lambda_{\odot}= 200.897^{\circ} \pm 0.005^{\circ}$ for optical detections reported by \cite{Jenniskens2020ACari}. We further elaborate on SAAMER-OS' collecting area and meteoroid flux estimates in \ref{sec:flux}. The geocentric ecliptic radiant positions, geocentric speed, and orbital elements derived with the wavelet analysis are listed in Table \ref{table}. We employ 10,000 iterations in a Monte Carlo procedure to draw orbital element uncertainties from the errors in radiant position and speed. We follow \cite{Bruzzone2020} to estimate the error in the outburst radiant position as the angular separation of the wavelet radiant position and the position of peak radiant density at $\lambda_{\odot}=201^{\circ}$. On the same date, we adopt the error in the outburst speed as the standard error of the outburst mean geocentric speed. The errors in the outburst radiant position and speed are $0.7^{\circ}$ and $0.09$ km s$^{-1}$ respectively. The orbital elements derived with SAAMER-OS closely resemble those obtained by \cite{Jenniskens2020ACari} with video observations. Employing the $D^{\prime}$ criterion \citep{Drummond1981} to look for potential parents for this outburst from the Minor Planet Center Orbit Database\footnote{\href{http://www.minorplanetcenter.net/iau/MPCORB.html}{http://www.minorplanetcenter.net/iau/MPCORB.html} , retrieved October 20, 2020.} results in no clear parent object of the $\alpha$ Carinid meteor shower.

%
%\begin{table}[!h]
%\caption{Geocentric radiants (J2000.0) and orbits of the $\alpha$ Carinid.}\label{table_orbs}
%\begin{adjustwidth}{-4.0cm}{}
%\begin{tabular}{ccccccccccccccc}
%\hline
%\hline
%Reference & $\alpha_{g}$(deg) & $\delta_{g}$ (deg) & v$_{g}$ (km s$^{-1}$) & a (au) & q (au) & e & i (deg) & $\omega$ %(deg) & $\Omega$ (deg) & $N$\\
%\hline
%This work & 98.905 & $-53.671$ & $33.3 \pm0.1$ & $3.5\pm0.3$ & $0.9972\pm0.0001$ & $0.72\pm0.02$ & $55.8\pm0.3$ & %1$\pm$173 & 21.7 & 1352\\ 
%\cite{Jenniskens2020ACari} & 98.7 & $-54.3$ & $32.4\pm1.4$ & 3.31 & $0.977\pm0.004$ & $0.696\pm0.09$ & $54\pm2$ & %$0.8\pm1.9$ & $20.9\pm0.1$ & 130\\  
%\cite{Jenniskens2018}& 103.2 & $-57.0$ & 30.1 & 2.87 & 0.989& 0.655 & 50.4 & 354.7 & 18 & 121\\  
%\hline
%\end{tabular}
%\end{adjustwidth}
%\end{table}
In addition to the outburst reported here, two more sudden outbursts have been detected with SAAMER-OS south of the South Toroidal region: the Volantids outburst on December 31, 2015 \citep{Bruzzone2020}: $\lambda_{\odot}=280^{\circ}$, $\lambda_{g}-\lambda_{\odot}=304^{\circ}.1$, $\beta_{g}=-77.7$, $v_{g}=30.2$ km s$^{-1}$, and the $\beta$ Tucanid / $\delta$ Mensid in March 12, 2020 \citep{Janches2020outburst}, $\lambda_{\odot}=352^{\circ}$, $\lambda_{g}-\lambda_{\odot}=305.73^{\circ}.1$, $\beta_{g}=-77.2$, $v_{g}=30.7$ km s$^{-1}$. Radiant and orbital elements are listed in Table \ref{table}. All three outburst orbits resemble those of a short period comet, but display similarly higher inclinations than expected for Jupiter-family comets. The orbits share similar shape, size, and inclination; however their orientations differ and both $\omega$ and $\Omega$ display more scatter. The orbits are close to several important Mean Motion Resonances (MMR) with Jupiter, especially to the 2:1, 5:3, 8:5 and 7:3 at $e=0.7$ and $i=55^{\circ}$ \citep{Gallardo2020} and differ from those at the South and North Toroidal regions: $a\sim 1$ au, $e \sim 0.2$, $i\sim 60^{\circ}-70^{\circ}$ \citep{Campbell-Brown2008, Janches2015}. Furthermore, the duration of these outburst may indicate they are part of relatively younger streams as opposed to the older sporadic meteoroids comprising the Toroidal ring which likely evolved from long period comet-type objects \citep{Pokorny2014}. We note however, that the duration for these outbursts (between two and seven days) suggest fairly evolved streams as opposed to very young ones such as the Camelopardalids. \citealt{Janches2020outburst} report asteroid (248590) 2006 CS ($a=2.91$ au, $e=0.7$, $i=52^{\circ}.3$, $\Omega=172^{\circ}.4$, $\omega=346^{\circ}.4$) as a promising parent candidate ($D^{\prime}=0.055$) of the $\beta$ Tucanid / $\delta$ Mensid shower outburst. The latter may suggest that the three detected outbursts could have originated from a short-period object. However, further analysis including dynamical simulations will be needed to properly address the origin of these outbursts.
\begin{table}[!h]
\caption{Meteor shower outburst observed with SAAMER-OS.}\label{table}
\begin{adjustwidth}{-3.2cm}{}
\begin{tabular}{cccccccccc}
\hline
\hline
Object & $\lambda_{g}-\lambda_{\odot}$ (deg) &$ \beta_{g}$ (deg) &$v_{g}$ (km s$^{-1}$) & a(au)&q(au)&e&i(deg)& $\omega$ (deg) & $\Omega$ (deg)\\
\hline
$\alpha$ Carinid &  $271.04$ & $-76.4$ & 33.3 &  $3.5\pm0.3$ & $0.9972\pm0.0001$ & $0.72\pm0.02$ & $55.8\pm0.3$ & 1$\pm$173 & 21.7 \\
$\beta$ Tucanid /$\delta$ Mensid & $305.7$ & $-77.2$ & $30.7$ & $3.2\pm0.1$& $0.976\pm0.001$ & $0.69\pm0.01$ & $50.8\pm 0.2$ & $345.3\pm0.5$& 172.0 \\
VOL & 304.1 & $-77.7$ & 30.2 &  $3.1\pm0.7$ &$ 0.970\pm0.001$ & $0.69\pm0.04$& $49.7\pm0.8$&$166.2\pm0.7$ & 280.0 \\
\hline
\end{tabular}
\end{adjustwidth}
\end{table}

\subsection{Estimating SAAMER-OS' collecting area and meteoroid fluxes}\label{sec:flux}
In order to estimate meteoroid fluxes with SAAMER-OS, a measure of the radar collecting area and a correction for observational biases is needed. The meteoroid flux $\Phi$, can be estimated by dividing the debiased meteor rate $\Sigma$ by the radar collecting area $A$ \citep{Campbell-Brown2004, Campbell-Brown2015, Bruzzone2015}. Biases affecting radar observations are numerous and pertain to the specific radar system parameters, the interaction of the scattered waves within the atmosphere, and the inherent scattering mechanism of radio waves by free electrons \citep[cf.][]{Ceplecha1998, Galligan2004, Galligan2005}. Such biases include the initial trail radius effect, in which the observability of meteors occurring higher in the atmosphere is reduced due to the increase of the mean free path with height, resulting in the attenuation of the echo amplitude for large trail widths due to destructive interference. Other effects include Faraday rotation, the change of the polarization plane of the radar wave as it passes through the ionosphere; the diffusion of meteor trails during formation and the decay time of established meteor trails. The combination of these effects result in a decrease in the meteor rates for any radar system. To correct for observational biases we make use of the derivation of SAAMER-OS' RRF in \cite{Janches2015} and refer the reader to that study for an in-depth description and derivation of correction factors for this system. We model the transmitting and receiving antennas with a NEC-2D code to determine the beam patterns. We find that the total gain and beam patterns for the antennas are the same and well described by a smooth function on elevation alone and proceed to fit it with a nine-order polynomial. The peak gain is 8.8 dB at the zenith ($z=0^{\circ}$) with the -3 dB point at $z=79^{\circ}$. Correction for Faraday rotation is not necessary since SAAMER-OS receiving antennas are cross-Yagis and thus the system receives both linear polarizations and is not sensitive to this effect \citep{Janches2015}. % Once all meteor detections are corrected with the RRF, we proceed to estimate the radar collecting area and to determine meteoroid fluxes. 
To estimate the radar collecting area,  we use the methodology in \cite{Kaiser1960,Brown1995} and \cite{Brown1998} as a guide. The collecting area is a strip of space that is perpendicular to the meteor radiant and has a width given by the mean vertical trail length, which describes the altitude range in which ablation occurs. The length of the strip is the length of the echo line which extends from horizon to horizon. In practice, the echo line is truncated out to a limiting range for the radar. For the vertical trail height, we use the empirical relation with mass index $s$ reported in \cite{Brown1998} from TV observations of faint meteors by \cite{Flemming1993}. Values of $s$ below 2.0 indicate that there is more mass in larger particles, where the opposite holds for values larger. In general, values for showers are in the $1.6~-2.0$ range whereas values for sporadics are larger than 2.0 \citep{Blaauw2011shower,Blaauw2011sporadic}. We adopt $s=2.0$ for the flux estimates in this work. However, we also include fluxes for a list of $s$ values and leave the development of a method to determine specific shower mass indices with SAAMER-OS for a future study. 

We parameterize the echo line as a function of its elevation $\phi$, weighted by the antenna gain, and obtain the length through numeric integration. Since the vertical  trail length is only dependent on $s$, the collecting area $A(z,h,s)$ can be found by multiplying the echo line length by the vertical trail length. Following \cite{Campbell-Brown2004, Campbell-Brown2015} and \cite{Bruzzone2015}, we set $h=100$ km. 
After some algebra \citep[see][]{CBaumann2012}, the collecting area $A(z,h,s)$ can be expressed as:
\begin{equation}
A(z,h,s)=2\int_{0}^{\pi/2}\left(G(\phi)\cos z\right)^{s-1}R_{E}\cos \phi  \left(\frac{\frac{R_{E}}{R_{E}+h}\sin \phi} {\sqrt{1-\left(\frac{R_{E}}{R_{E}+h}\cos \phi  \right)^{2}}}-1\right)\frac{d\phi}{\sin z}\left(1.15+14.6e^{\frac{-s}{1.44}}\right),
\end{equation}
where $G(\phi)=\sqrt{G_{\text{Tx}}(\phi)G_{\text{Rx}}(\phi)}$ is the radar gain pattern, $z$ the radiant zenith distance, $R_{E}$ is Earth's radius, $h$ is the meteor height, $s$ is the mass index, and $\phi$ the echo line elevation. Here $G_{\text{Tx}}$ and $G_{\text{Rx}}$ are the radar transmit and receive antenna gain powers, which in this case are identical. 

\begin{figure}[ht!]
 \centering
 \includegraphics[width=4.0in]{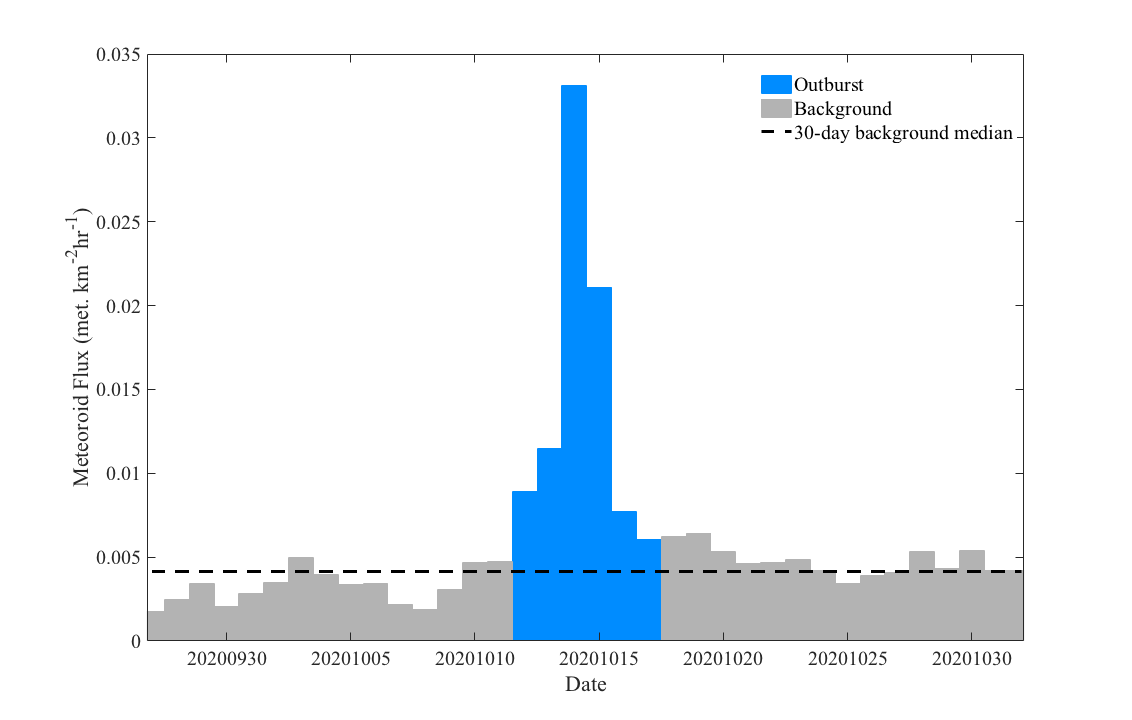}
  \caption{$\alpha$ Carinid 24hr-average fluxes with SAAMER-OS with $s=2.0$. Fluxes associated with the shower by the wavelet-based transform displayed in blue with background fluxes in grey. The dashed line marks the 30 day-median background flux of 0.00415 met. km$^{-2}$ hr$^{-1}$}\label{24hrflux}
\end{figure}

Outburst fluxes are estimated at one hour intervals within a $6^{\circ}\times6^{\circ}$ window centered on the position of wavelet-based outburst radiants for each day from October 12 through October 17 2020. Windows are partitioned in $0.1^{\circ}$ steps and the debiased meteor rates and collecting area computed. Hourly fluxes are then determined by finding the ratio $\Sigma:A$. Hourly $\alpha$ Carinid fluxes in Figure \ref{fig_profile} are determined as the total sum of fluxes in the window within $3^{\circ}$ from the position of the wavelet radiants. The shower daily flux is computed from the averaged hourly fluxes where we subtract the equivalent sporadic background averaged over 15 consecutive days before October 12, and 15 days after October 17, 2020. To help the comparison with results from visual observers, we follow  \cite{Koschack1990} and adjust the flux to a +6.5 limiting magnitude, $\Phi_{+6.5}$, using
\begin{equation}
\Phi_{+6.5}=\Phi_{+9.0}\times10^{\left(6.5-9.0\right)\left(s-1\right)/2.5},
\end{equation}
and estimate the zenithal hourly rates (ZHR) with
\begin{equation}
ZHR(r)=37200~ \text{km$^{2}$} \times \Phi_{+6.5}\times\left(\frac{\left(r-1.3\right)^{-0.748}}{13.1r-16.45}\right),
\end{equation}
where $r=10^{\left(s-1\right)/2.5}$ is the population index. Figure \ref{24hrflux} displays the average meteoroid fluxes for the outburst, before sporadic flux subtraction, and sporadic background fluxes. The dashed line indicates the 30 day-median sporadic flux. After background subtraction, we record a peak average flux, $\Phi_{+9.0}=0.029$ meteoroids km$^{-2}$ hr$^{-1}$, down to a +9 limiting magnitude, on October 14, 2020. The peak flux corresponds to a ZHR$_{\text{max}}$ of approximately 5.7 at $r=2.5$. The peak ZHR is comparable to values for other established showers from optical observations at similar $r$ values: Southern and Northern Taurids (ZHR$_{\text{max}}=5$, $r=2.3$), September $\epsilon$ Perseids (ZHR$_{\text{max}}=5$, $r=2.5$), Aurigids (ZHR$_{\text{max}}=6$, $r=2.5$), $\alpha$ Capricornids (ZHR$_{\text{max}}=4$, $r=2.5$) and $\gamma$ Normids (ZHR$_{\text{max}}=4$, $r=2.4$) \citep{Rendtel2014}.
Instead of displaying a fixed $s$ value, in general the shower mass index drops as the Earth intersects the core of the stream where larger particles are located \citep{Blaauw2011shower}. For this reason, we include peak flux estimates for a list of mass index values typical for showers in Table \ref{table_flux}. The wide range in ZHR values reflects the sensitivity of shower fluxes with mass index. Our flux estimates suggest that the outburst flux likely never rises above the level seen in showers like the $\eta-$Aquariids, ZHR$_{\text{max}}=50-80$, adopting $s=1.9$ \citep{Campbell-Brown2015}, or the Daytime Arietids, ZHR$_{\text{max}}=189$ at $s=1.75$ \citep{Bruzzone2015}.

We revisit archival observations of the previous $\beta$ Tucanid shower outburst detected with SAAMER-OS on March 12, 2020 \citep{Janches2020outburst}, and apply the procedure developed here to estimate the average peak flux. We find a maximum 9-magnitude flux for the $\beta$ Tucanid shower of 0.01 met. km$^{-2}$ hr$^{-1}$ at $s=2.0$ which corresponds to a ZHR$_{\text{max}}$ slightly above 2.

\begin{table}[!h]
\begin{center}
\caption{$\alpha$ Carinid fluxes for various mass index values following the procedure outlined in Section \ref{sec:flux}.}\label{table_flux}
\begin{tabular}{cccc}
\hline
$s$ & $\Phi_{+9}$ & $\Phi_{+6.5}$ & ZHR\\
\hline
  1.65     &  2.24e-2  &  5.33e-3  &    43.8\\
  1.70     &  2.25e-2  &  4.89e-3  &    31.1\\
  1.75     &  2.52e-2  &  4.50e-3   &   22.7\\
  1.80     &  2.60e-2  &  4.12e-3   &   16.8\\
  1.85     &  2.68e-2  &  3.78e-3  &    12.6\\
  1.90     &  2.75e-2  &  3.47e-3   &   9.6\\
  1.95     &  2.83e-2  &  3.17e-3  &    7.4\\
  2.00     &  2.90e-2  &  2.90e-3   &   5.7\\
  2.05     &  3.00e-2  &  2.66e-3  &    4.4\\
  2.10     &  3.04e-2  &  2.43e-3   &   3.5\\
  2.15     &  3.11e-2  &  2.23e-3   &   2.7\\
  2.20     &  3.18e-2  &  2.04e-3   &   2.2\\
\hline
\end{tabular}
\end{center}
\end{table}

\section{Conclusions}\label{sec:conclusions}
We reported radar observations of an unexpected outburst of the $\alpha$ Carinid meteor shower recorded with the Southern Argentina Agile MEteor Radar Orbital System (SAAMER-OS). Our wavelet-based analysis returned the shower radiant location south of the South Toroidal ring at $\lambda_{g}-\lambda_{\odot}=271^{\circ}.04$,  $\beta_{g}=-76^{\circ}.4$ with $v_{g}=33.3$ km s$^{-1}$ during the peak at $\lambda_{\odot}=201^{\circ}$ on October 14, 2020. The wavelet-based technique unequivocally confirms the sudden appearance of the outburst in 2020 rising 16 times the total standard deviation above the annual median. The outburst lasted for approximately 6 days starting on October 12 through October 17, 2020. The radiant location, speed and period of observation agree with those reported with video observations by \cite{Jenniskens2020ACari}. We measured a 9-magnitude peak hourly flux of 0.09 meteoroids km$^{-2}$ hr$^{-1}$, assuming $s=2$, down to a limiting mass of $1.9\times10^{-8}$ kg between 21 UTC and 22 UTC on October 14. To compute fluxes we debiased the observed meteor rates and estimated the radar collecting area at one hour intervals. The 6 magnitude-equivalent peak average daily flux corresponds to a ZHR of approximately 6, comparable to other known meteor showers at similar $s$ values. We computed peak average fluxes for several mass index values to derive probable ZHR estimates returning limits between 2 and 44 approximately. The latter suggest that $\alpha$ Carinid fluxes remain well below those recorded for strong showers like the Daytime Arietids or Geminids. Based on 1352 events during the peak, the orbital elements resemble those of a short-period object: $a=3.5\pm0.3$ au, $q\simeq 1$ au, $e=0.72\pm 0.02$, $i=55^{\circ}.8\pm0^{\circ}.3$, $\omega=1^{\circ}\pm 173^{\circ}$, $\Omega=21^{\circ}.7$. Comparably, two other austral shower outbursts previously recorded with SAAMER-OS, the $\beta$ Tucanid / $\delta$ Mensid and the Volantids (VOL), have orbits with shape, size and inclination similar to the $\alpha$ Carinid. Our search for a parent object using the $D^{\prime}$ criterion \citep{Drummond1981} did not reveal any clear candidate. While the duration suggests the shower is not as old as the Arietids or Taurid streams (thousands to tens of thousands of years), the significant spread in nodal crossing may indicate a fairly evolved stream.
\section{Acknowledgements}
J.S.B. and D.J.'s work is supported by the NASA SSO and ISFM Programs and the NASA NESC. RJW was supported through NASA grant 80NSSC18K0656. SAAMER's operation is supported by NASA SSO, NESC assessment TI-17-01204, and NSF grant AGS-1647354. The authors appreciate the invaluable support of Carlos Ferrer, Gerardo Connon, Luis Barbero and Leandro Maslov with the operation of SAAMER.
\bibliography{references}
\end{document}